\documentclass{amsart}

\usepackage{amsmath}
\usepackage{amsthm}
\usepackage{amsfonts}
\usepackage{amssymb}
\usepackage{graphicx}
\usepackage{xcolor}
\usepackage{setspace}
\usepackage{subcaption}
\usepackage{etoolbox}
\usepackage[foot]{amsaddr}
\usepackage{etoolbox}
\usepackage{tikz}
\usepackage{colortbl}
\usepackage{makecell}
\usepackage{stmaryrd}

\usepackage[authoryear,compress]{natbib}

\usepackage{hyperref}
\hypersetup{
    colorlinks,
    linkcolor={red!50!black},
    citecolor={blue!50!black},
    urlcolor={blue!80!black}
}

\makeatletter
    \patchcmd{\NAT@test}{\else \NAT@nm}{\else \NAT@nmfmt{\NAT@nm}}{}{}
    \DeclareRobustCommand\citepos
        {\begingroup
        \let\NAT@nmfmt\NAT@posfmt
        \NAT@swafalse\let\NAT@ctype\z@\NAT@partrue
        \@ifstar{\NAT@fulltrue\NAT@citetp}{\NAT@fullfalse\NAT@citetp}}
    \let\NAT@orig@nmfmt\NAT@nmfmt
    \def\NAT@posfmt#1{\NAT@orig@nmfmt{#1's}}
\makeatother

\title[Moral Dilemmas for Moral Machines]{Moral Dilemmas for Moral Machines}

\author{Travis LaCroix}

\address{Department of Philosophy Dalhousie University}

\email{tlacroix@dal.ca}

\date{Draft of \today. Please cite official version, \href{https://doi.org/10.1007/s43681-022-00134-y}{https://doi.org/10.1007/s43681-022-00134-y}}

\begin{document}

\maketitle

\begin{abstract}
\singlespacing
    
    \phantom{a} Autonomous systems are being developed and deployed in situations that may require some degree of ethical decision-making ability. As a result, research in machine ethics has proliferated in recent years. This work has  included using moral dilemmas as validation mechanisms for implementing decision-making algorithms in ethically-loaded situations. Using trolley-style problems in the context of autonomous vehicles as a case study, I argue (1) that this is a misapplication of philosophical thought experiments because (2) it fails to appreciate the purpose of moral dilemmas, and (3) this has potentially catastrophic consequences; however, (4) there are uses of moral dilemmas in machine ethics that are appropriate and the novel situations that arise in a machine-learning context can shed some light on philosophical work in ethics.
    
    \phantom{a}

\end{abstract}

\setcounter{page}{1}

\section{Introduction}
\label{sec:Introduction}

    Increasingly, autonomous systems are being developed and deployed in situations that may require some degree of ethical decision-making ability. Some well-discussed examples include autonomous weapons for warfare \citep{Arkin-2008a, Arkin-2008b, Krishnan-2009, Tonkens-2012, Hellstrom-2013, Asaro-2020}; professional service robots for healthcare and elderly care \citep{Anderson-et-al-2006, Anderson-Anderson-2008, Sharkey-Sharkey-2012, Conti-et-al-2017}; sex robots for therapy or personal pleasure \citep{Eichenberg-et-al-2019, Headleand-2020, Doring-et-al-2020}; and self-driving vehicles for transportation \citep{Bhargava-Kim-2017, Sommaggio-Marchiori-2018, Evans-et-al-2020}.
    
    In the early days of machine learning (ML), researchers could focus on the fundamental aspects of their work without much concern for social or ethical consequences since these systems were relatively encapsulated within the confines of the research lab. However, \citet{Luccioni-Bengio-2019} highlight that these algorithms are increasingly being deployed in society. This is due, in part, to the promise of the unprecedented economic impacts of ML applications \citep{Bughin-et-al-2018, Szczepanski-2019, Russell-2019}. As a result, research in machine ethics---including fundamental questions surrounding the very nature and possibility of artificial moral agency---has proliferated in recent years.\footnote{So too have attempts to codify principles for ethical AI research, though largely to little effect. See \citet{Jobin-et-al-2019} for a recent survey; see also \citet{LaCroix-Mosheni-2021} for a discussion of the efficacy of such proposals.} This work has included  using moral dilemmas (i.e., philosophical thought experiments) as validation mechanisms for  implementing decision-making algorithms in ethically-loaded situations.  
    
    This paper aims to describe the use of philosophical thought experiments in the context of machine ethics research and explain how these experiments are misused in this field. As I argue, this misapplication comes from an apparent misunderstanding of what morally charged thought experiments from philosophy are supposed to accomplish. I conclude by describing what philosophical thought experiments are useful for in the context of ML and addressing some meta-ethical worries. I also describe how the novel situations that arise from the possibility of autonomous agents can shed some further light on philosophical work in ethics.  
    
    As a concrete example, I will focus on trolley-style problems applied to hypothetical scenarios that may be faced by autonomous vehicles since this is perhaps the most prevalent (mis-)use of a philosophical thought experiment in the context of artificial intelligence systems; however, I will also gesture toward other examples when applicable, to not give the (false) impression that this is a relatively isolated case.

\section{Moral Machines}

In this section, I discuss how philosophical thought experiments---particularly, moral dilemmas---are used in machine learning to benchmark the `ethical' performance of new algorithms. As a case study, I begin by providing some technical background on how autonomous vehicles work, which makes salient several possible problems arising in situations where a decision may need to be rendered that has potential moral weight (\ref{sec:AVs}). This situation gives rise to the appropriation of trolley-style problems in the context of autonomous vehicles needing to `choose' between two undesirable alternatives (\ref{sec:Trolley-Style}). I then describe a well-known use case of trolley-style problems in machine ethics: the {\it Moral Machine Experiment} (\ref{sec:MoralMachines}). This section concludes with a discussion of moral dilemmas more generally, highlighting how they are taken in the machine learning literature as {\it benchmarks} for determining whether an algorithm `acts ethically' (\ref{sec:Benchmarking}). In the subsequent section, I argue that this view is mistaken.

\subsection{Autonomous Vehicles}
\label{sec:AVs}

    The Society for Automotive Engineers defines six levels of automation, ranging from (0) no automation, where the driver performs all driving tasks, to high (4) or full (5) automation, where the vehicle is capable of autonomously performing all driving functions under certain/all conditions. Most vehicles on the road today are classed as level $0$ or $1$: they are controlled by humans but may have some driver-assistance capabilities, such as adaptive cruise control. However, several vehicles on the market from several different manufacturers fall under level $2$ or $3$---partial and conditional automation, respectively. For example, Tesla's level-$2$ autopilot function partially automates the vehicle, but a human driver's attention is still legally required at all times. Honda was the first company to have a vehicle classed with the level-3 designation, although this model has yet to be mass-produced---as of September 2021,  the public sale of this vehicle was limited to 100 units in Japan. 
    Predictions vary widely as to when fully-autonomous vehicles will be available for private use, which is consistent with the long history of overestimating the near-future abilities of AI systems.%
        \footnote{For example, the goal of the Dartmouth Summer Research Project on Artificial Intelligence, held in 1956 and organised by John McCarthy, Marvin Minsky, Nathaniel Rochester, and Claude Shannon, was stated as follows: 
            \begin{quote}
                The study is to proceed on the basis of the conjecture that every aspect of learning or any other feature of intelligence can in principle be so precisely described that a machine can be made to simulate it. An attempt will be made to find how to make machines use language, form abstractions and concepts, solve kinds of problems now reserved for humans, and improve themselves. We think that a significant advance can be made in one or more of these problems if a carefully selected group of scientists work on it together for a summer.
            \end{quote}
        Of course, these are all still open problems today \citep{Russell-2019}.} 

    Machine learning algorithms for autonomous vehicles must continuously render the surrounding environment, in addition to predicting possible changes to that environment moving forward through time and reacting appropriately to those changes. This ability involves a suite of applications, including object detection, recognition, localisation, and prediction (of movement). For example, many autonomous vehicles utilise RADAR (radio detection and ranging) and LIDAR (light detection and ranging) sensors, in addition to video cameras (to measure distance, detect road edges, and identify lane markings) and ultrasonic sensors (to detect curbs and other vehicles). These data may be fed into (typically several) deep neural networks and processed in real-time.%
        \footnote{The individual tasks---detection, recognition, localisation, prediction, action---can be accomplished using several different methods---including regression, pattern recognition, clustering, and decision matrices, often involving a plethora of state-of-the-art machine-learning techniques. For example, support vector machines and principal component analysis may be used for pattern recognition; K-means clustering may be used to identify data in low-resolution images; and gradient-boosting may be used for decision-making, depending on confidence levels for detection, classification, and prediction. Advances are continually being made in this field, with some focus on end-to-end learning; for example, \citet{Bojarski-et-al-2016, Yang-et-al-2017} employ convolutional neural networks to train their vehicles without requiring the highly complex suite of algorithms used in traditional methods. \citet{Kuefler-et-al-2017} use Generative Adversarial Networks to train their system by mimicking human behaviour.}

    The advent of artificial intelligence systems highlights how difficult it is to perform tasks that humans take for granted. For example, the sensors of an AV must detect when a pedestrian is waiting at a crosswalk, recognise that it is a pedestrian, predict whether the pedestrian will step out into the road, and respond appropriately---i.e., by slowing down or stopping. And, strange things can happen when the system is presented with examples that it has not yet encountered: in 2018, a self-driving vehicle in Tempe, AZ apparently alternated between classifying a pedestrian, Elaine Herzberg, who was walking her bicycle in the street, as `vehicle', `person', and `other object'. The result was that the vehicle struck and killed Herzberg \citep{Wakabayashi-2018}.
    

    Part of the difficulty arises from the training data. In Herzberg's case, the system could not recognise a human walking with a bike as two separate things needing classification (among many other things that went wrong). In addition, the software used in this particular instance did not include considerations for jaywalking pedestrians. If training examples always include crosswalks, the system may pick up on these underlying regularities instead of the intended target. 
    For example, suppose an image-classification algorithm has only ever seen red apples. In that case, it might misclassify a green apple as a pear because its `concept' of {\sc apple} depends (too) heavily on some spurious regularities in the examples it has seen \citep{Christian-2020}.

    Let's suppose that these problems are surmountable and fully autonomous vehicles are achievable in the foreseeable future.%
        \footnote{In fact, full automation is not necessary for the problems described to arise since human reaction time will not be useful in split-second moral decisions.} 
    As autonomous vehicles become more prevalent on society's roads, it is supposed that it will become increasingly likely that an individual vehicle will need to be programmed to make decisions in situations that carry significant moral weight. Practically, this is a difficult problem. Crashes are relatively rare in terms of the data that a machine-learning system might receive; therefore, the system may not `know' how to respond, because of a lack of training data. Low-probability, but high-{\it risk}, events pose particular challenges for machine learning methods that depend upon the system seeing many examples in order to learn. This is true even when there is an objectively correct answer to the problem; however, in morally-charged situations, there may not be obviously correct answers on which to train the model, as I will discuss below.
    

\subsection{Trolley-Style Problems}
\label{sec:Trolley-Style}

    In some (perhaps exceedingly rare) circumstances, an autonomous vehicle may face a situation that can be classified as a trolley-style problem. As is well-known in philosophy, the trolley problem is a set of ethical dilemmas wherein a subject must choose between some set of options involving (typically) human lives. The problem was first introduced in \citet{Foot-1967} in a discussion of abortion and the doctrine of double-effect. This initial problem was expanded upon and later analysed in much more detail by \citet{Thomson-1976} and \citet{Unger-1996}. I refer to these as trolley-{\it style} problems because, although they have the same basic structure as a trolley problem, their scope has been expanded well beyond the original philosophical context.
        
    Despite the rarity of trolley-style problems in the real world, the extent to which they are possible implies that the machine will need to `know' how to react. As a result, the advent of autonomous vehicles has re-invigorated interest within and without philosophy on the subject of trolley problems; however, as noted in the introduction, this is but one salient example of a more general interest in moral dilemmas.%
        \footnote{In the last few years alone, there have been dozens of articles that refer to Philippa Foot's 1967 paper in the context of autonomous vehicles; see, for example, \citet{Allen-et-al-2011, Wallach-Allen, Pereira-Saptawijaya-2015, Pereira-Saptawijaya-2011, Berreby-et-al-2015, Danielson-2015, Lin-2015, Malle-et-al-2015, Saptawijaya-Pereira-2015, Saptawijaya-Pereira-2016, Bentzen-2016, Bhargava-Kim-2017, Casey-2017, Cointe-et-al-2017, Greene-2017, Lindner-et-al-2017, Santoni-de-Sio-2017, Welsh-2017, Wintersberger-2017, Bjorgen-et-al-2018, Grinbaum-2018, Misselhorn-2018, Seoane-Pardo-2018, Sommaggio-Marchiori-2018, Baum-et-al-2019, Cunneen-2019, Krylov-et-al-2019, Sans-Casacuberta-2019, Wright-2019, Agrawal-et-al-2020, Awad-et-al-2020, Banks-2020, Bauer-2020, Etienne-2020, Gordon-2020, Harris-2020, Lindner-et-al-2020, Nallur-2020}. And, several more articles that discuss trolley problems without citing Foot; e.g., \citet{Bonnefon-et-al-2016, Etzioni-Etzioni-2017, Lim-2019, Evans-et-al-2020}; or, which appear to reinvent the trolley problem (without citing Foot); e.g., \citet{Keeling-2018}.}
    
    We suppose that an autonomous vehicle is about to crash and has no trajectory to save everyone. Is it better, for example, to hit a group of pedestrians on the road or swerve into a barrier, killing the driver? When harm is possible or inevitable, the vehicle will need to make a decision, which means that it needs to have been programmed or trained to be capable of making a decision. And, this is true regardless of how rare the circumstances might be in practice. In response to these facts, \citet{Awad-et-al-2018} have noted that it will be necessary to gauge social expectations about how to divide the risk of harm between the different stakeholders on the road. Their response to this is the {\it Moral Machine Experiment}.

\subsection{The Moral Machine Experiment}
\label{sec:MoralMachines}

    The Moral Machine Experiment \citep{Awad-et-al-2018} is a multilingual online `game' for gathering human perspectives on moral dilemmas---specifically, trolley-style problems in the context of autonomous vehicles. By the time of publication, the Moral Machine Experiment had collected nearly $40$ million data points from around the world. Individuals who participated were shown many unavoidable accident scenarios with binary outcomes---i.e., trolley-style problems---and were prompted to choose the scenario they prefer. These include `sparing humans (versus pets), staying on course (versus swerving), sparing passengers (versus pedestrians), sparing more lives (versus fewer lives), sparing men (versus women), sparing the young (versus the elderly), sparing pedestrians who cross legally (versus jaywalking), sparing the fit (versus the less fit), and sparing those with higher social status (versus lower social status)' (60). Some scenarios include other `characters', such as criminals, pregnant women, or doctors.%
        \footnote{There are obvious and perhaps pressing philosophical questions that arise concerning some of these classifications, but we will put those aside for now.} 
    Globally, they find that individuals tend to prioritise humans over animals, many humans over fewer, and younger humans over older.

    According to Edmond Awad---one of the paper's coauthors---the original purpose of the Moral Machine Experiment was supposed to be purely descriptive, highlighting {\it people's} preferences in ethical decisions \citep{Vincent-2018}. However, the first, second, and last authors (Awad, Dsouza, Rahwan) of the original paper, citing the results of the {\it Moral Machine Experiment}, published an article a month later which proposes a `voting-based system for ethical decision making' \citep{Noothigattu-et-al-2018}. They suggest that the Moral Machine Experiment data can be used to automate decisions, `even in the absence of such ground-truth principles, by aggregating people’s opinions on ethical dilemmas' (4). This statement takes the descriptive project into the realm of normative ethics by suggesting that the Moral Machine Experiment data can serve as a validation mechanism for whether an algorithm acts `morally'.

\subsection{Benchmarking Ethical Decisions}
\label{sec:Benchmarking}

    Several authors have proposed algorithms for moral decision making in autonomous vehicles, and there are intrinsic reasons why we might want AI systems to be capable of acting ethically. However, for-profit corporations have additional incentives for designing `ethical' AI since humans (i.e., {\it consumers}) will likely be more trusting of an autonomous agent (i.e., {\it products}) if it is known to possess a set of moral principles intended to constrain and guide its behaviour \citep{Bonnefon-et-al-2016}. The question then arises how we are supposed to know whether the decision chosen by the system is `in fact' moral or not---i.e., {\it how} ethical are the decisions made by the algorithm?
    
    Benchmarking is a way of evaluating and comparing new methods in ML for performance on a particular dataset \citep{Olson-et-al-2017}. Following \citet{Raji-et-al-2021}, we can understand a {\it benchmark}, for the purposes of this paper, as a dataset plus a metric for measuring the performance of a particular model on a specific task. For example, suppose that the current state of the art of image classification on ImageNet for top-1 accuracy is 85\%. In this case, 85\% top-1 accuracy is a benchmark---namely, an objective measure of how well one's algorithm performs on a particular dataset for image classification. So, if a method performs worse than this benchmark, we know that something has gone awry. However, if the new method performs {\it better} than this benchmark, it is the best-performing algorithm to date (again, modulo efficiency, the volume of training data, etc.). By and large, state-of-the-art progress on certain benchmarks is typically taken to indicate progress on a particular task or set of tasks \citep{Raji-et-al-2021}.

    Moral dilemmas have been appealed to as benchmarks for checking whether an algorithm makes the `right' decision in machine ethics. The most popular method for {\it evaluating} whether an artificial system behaves ethically is by evaluating its performance on ethical dilemmas \citep{Nallur-2020}. In the case of autonomous vehicles, the most common dilemma that is appealed to is the trolley problem---due, in no small part, to its use in the Moral Machine Experiment. Using the Moral Machine Experiment as a benchmark can be understood in the following way. The dataset is the survey data that was collected by the experimenters---i.e., which of the binary outcomes is preferred by participants, on average. If human agents strongly prefer sparing more lives to fewer, then researchers might conclude that the `right' decision for their algorithm to make is the one that reflects this sociological fact. Thus, the metric would measure how close the algorithm's decision is to the aggregate survey data.\footnote{Of course, this approach raises all the usual problems of biased data, insofar as certain individuals are going to be overrepresented---i.e., those individuals from countries who have easy access to the internet. \citet{Falbo-LaCroix-2022} argue that these considerations may exacerbate structural inequalities and mechanisms of oppression---although, they also note that `more data' is not necessarily going to fix that, since the data are reflective of extant inequalities in society. However, it is crucial to note that the thing being measured in this case is {\it not} how ethical the decision is, but how closely the decision accords with the opinions of humans, on average.} Note, then, that {\it survey data} is being used as a proxy for {\it moral facts}.

        As previously mentioned, the notion of using moral dilemmas as a benchmark for machine ethics extends well beyond the particular case of trolley-style problems and autonomous vehicles. \citet{Lourie-et-al-2020} introduce a dataset of ethical dilemmas which they say `enables models to learn basic ethical understanding'. However, as with the Moral Machine Experiment, it is important to note that their metric for measuring performance can only measure how humans annotate the dilemma (i.e., whether a response is ethical or not) {\it on average}.

    In the more general case of moral dilemmas as benchmarks, \citet{Bonnemains-et-al-2018} explicitly reason as follows: since classic moral dilemmas have already been used as a basis for ethical reasoning, `it seems legitimate to use some of them as a starting point for designing an automated ethical judgement on decisions' (43). \citet{Bjorgen-et-al-2018} argue that {\it certain} types of ethical dilemmas `can be used as benchmarks for estimating the ethical performance of an autonomous system' (23). \citet{Kim-et-al-2018} construct a hierarchical Bayesian model for inferring individual and shared moral values from sparse and noisy data. Then they {\it evaluate} this approach by comparing their results with data from the Moral Machine Experiment. \citet{Cunneen-2019} suggest that the use of trolley-style problems as an elucidatory tool is a necessary {\it precedent} (i.e., is necessarily prior) to focusing AI applications on moral theories. And, \citet{Sutfield-et-al-2017} suggest that models of ethics (specifically for autonomous vehicles) should {\it aim} to match human decisions made in the same context.

    So, drawing normative consequences (an `ought') from some descriptive matters of fact (an `is') is not an unusual move in the field of machine ethics---particularly in the case of autonomous vehicles and trolley-style moral dilemmas. Besides being logically unsound, \citet{Etienne-2020} argues that this type of project is a dangerous basis for machine ethics, insofar as \citet{Awad-et-al-2018, Noothigattu-et-al-2018} lean on concepts of {\it social acceptability}, rather than, e.g., {\it fairness} or {\it rightness}; furthermore, individual opinions about these dilemmas may be highly volatile over time \citep{Henrich-et-al-2001, House-et-al-2013, Blake-et-al-2015}. 

    However, setting aside some of the logical and practical problems that using moral dilemmas as benchmarks entails, there is a sense in which these authors fail to appreciate the role of moral dilemmas in philosophical thought. In the next section, I provide a description of the {\it purpose} of philosophical thought experiments, including moral dilemmas, to show why using moral dilemmas for verification tools involves a category mistake.  In the conclusion, I discuss how this sets a dangerous precedent in the design of `ethical' AI systems.

\section{Philosophical Thought Experiments}

     As mentioned, moral dilemmas (a specific type of philosophical thought experiment), have been used as verification tools for moral decision-making in AI systems. In the case of the trolley-style problems posed by the Moral Machine Experiment (a specific type of moral dilemma), datasets---consisting of human responses to binary hypotheticals---and a measure of the system's predictive accuracy with respect to those responses (on average) are taken to provide a benchmark for determining how `ethically' the system acts. Here, I suggest that this is a mistake because, among other things, it misses the {\it point} of philosophical thought experiments.

    Thought experiments, historically, have served an important role in philosophical discourse---especially in the philosophy of science and metaphilosophy.%
        \footnote{There is a {\it vast} metaphilosophical literature on the role and purpose of thought experiments in philosophy. I do not have the space to delve into any adequate detail here; however, see \citet{sep-thought-experiment} for an overview. See also \citet{Asikainen-Hirvonen-2014} for a discussion of thought experiments in the context of science.} 
    According to \citet{Kuhn-1977}, thought experiments (generally) could make salient the failure of the world to conform to prior expectations about the way the world is. Similarly, thought experiments might elucidate particular ways in which a theory---i.e., one that is based on said prior expectations---might be revised to better conform with the facts about the world. 

    Following \citet{Brun-2018}, we can give the following simple schema for a thought experiment:
        \begin{enumerate}
            \item A scenario and a question are introduced;
            \item The experimenter elucidates the scenario and arrives at some result;
            \item A conclusion is drawn about some target(s). 
        \end{enumerate}
    A quick gloss on the {\it purpose} of thought experiments is that  they serve {\it primarily} as intuition pumps \citep{Dennett-1980, Dennett-1991, Dennett-2013}. They may be used, for example, to elicit normative intuitions, to justify counterfactual claims (also relying on intuitions), to explore logical relationships among philosophical theses, among others \citep{Mayo-Wilson-Zollman-2020-preprint}. As \citet{Dennett-1980} describes intuition pumps, they are typically used for {\it provoking} `a family of intuitions by producing variations on a basic thought experiment' (429). Importantly, philosophical thought experiments (as intuition pumps), should not be understood as `an engine of discovery, but a persuader or pedagogical tool---a way of getting people to see things your way' (429).

    So, moral dilemmas, like the trolley problem, provide some basic structure. A comparison of cases elucidating apparently incompatible or inconsistent reactions is supposed to shed light on some (morally) {\it salient} differences between the cases. This, in turn allows us to theorise about possible or plausible explanations for those differences. However, this too depends on individual intuitions to some extent: responses to moral dilemmas vary widely across societies and time periods \citep{Henrich-et-al-2001, House-et-al-2013, Blake-et-al-2015}. Trolley problems, specifically, have figured heavily in empirical research in neuroscience and psychology, and, again, human responses to these scenarios are highly dependent on external features.%
    \footnote{See, for example, \citet{Greene-et-al-2001, Greene-et-al-2004, Green-et-al-2008, Nichols-Mallon-2005, Cushman-et-al-2006, Schaich-et-al-2006, Ciaramelli-et-al-2007, Hauser-et-al-2007, Koenigs-et-al-2007, Waldmann-Dieterich-2007, Moore-et-al-2008}.} %

    A moral dilemma like the trolley problem pumps intuitions about why it may be permissible to perform an intentional action despite its foreseen and undesirable consequences. Intuitively, it seems like it may be permissible to pull a switch to divert the trolley from one track to another, as in the original trolley problem, despite that this would lead to the (foreseeable) death of one individual, because five people would be saved as a result.%
        \footnote{Surveys have shown that a vast majority would choose to pull the switch in this scenario~\cite{Navarrete-et-al-2012, Bourget-Chalmers-2014}.} %
    At the same time, it seems impermissible to actively kill someone---say, by pushing them on the track to block the trolley---despite that this would have an identical outcome: five lives saved at the expense of one.

    Thus, the dilemma gives rise to the question: Why might it {\it seem} morally permissible to act in one case but not in the other. Foot's proposed {\it explanation} for divergent intuitions in these two cases is that there is differential import between the {\it positive} and {\it negative} duties one has---in particular, negative rights (and the `duties' which follow from them) typically outweigh positive rights. In this case, the target of analysis is actually about the ethics of abortion---the ethical issue, it should go without saying, is not about trolleys. The thought experiment is useful because people are less likely to carry pre-theoretic baggage about trolleys than about abortions. Therefore, the trolley problem gets at the core of the issue in applied ethics while abstracting away from the moral loadedness of the actual target.

     This applies to philosophical thought experiments more generally, not just moral dilemmas. Consider a famous thought experiment: Gettier cases from epistemology \citep{Gettier-1963}. In a Gettier case, an individual clearly has a true and justified belief about a proposition, $p$, but it is not obvious that they {\it know} $p$. This is supposed to show that justified true belief cannot be sufficient for knowledge. But, for this thought experiment to work, it must be that in a Gettier case, the individual in question has a justified true belief that $p$ and {\it in fact} fails to know that $p$. However, knowledge is the very thing that epistemologists are trying to define---in this case, by deference to some set of (individually) necessary and (jointly) sufficient conditions. 
    
    Thus, if knowledge is a concept that requires analysis, the Gettier case cannot show that justified true belief is not knowledge since we do not know, {\it a priori}, what knowledge is. These examples depend upon the reader sharing Gettier's intuition that there is no knowledge in either of these cases. As such, the success of the Gettier cases is more of a sociological success than an epistemic one---what it shows is that {\it many philosophers share the intuition that there is no knowledge in Gettier cases}.%

    To make this point clearer, consider the following, perhaps unsatisfactory, possibility. Suppose epistemologists just {\it define} knowledge as justified true belief---so that these two concepts are functionally equivalent---and suppose that they are extremely rigid in this definition. If the entire epistemology community were to agree upon this definition of knowledge, then the Gettier case would not be a counterexample to a theory of knowledge {\it as} justified true belief because, in each instance, the individual has justified true belief and {\it therefore} (by definition) must have knowledge. The Gettier case is successful because of an intuition that is pumped in the vast majority of philosophers---i.e., that there {\it is not} knowledge in these instances. But, again, this says more about philosophers than it does about knowledge.


    Considering again what the trolley problem is supposed to show us, the conclusion to draw from is (obviously) not that it {\it is} moral to do $X$ in such-and-such a scenario and then to do $Y$ in the other (or vice-versa). Instead, the conclusion is best represented as a conditional statement; namely,
    \begin{quote}
        {\it IF} your intuitions are roughly such-and-such, \\ THEN $X$ {\it explains} why they are so. 
    \end{quote} %
    Where $X$ is filled in with some philosophical analysis. 
    
    However, what happens if intuitions diverge? Does this render Foot's analysis false? Or, is the divergent opinion false? Both of these questions are misguided. There is no `right' and `wrong' answer in a moral dilemma; there are only intuitions and explanations or theories about  the causes  of those intuitions.  To make the point in the most obvious possible way: a moral dilemma {\it is a dilemma}; it has no clear solution by design---or rather, it poses a problem that is inherently difficult, by design. Instead, moral thought experiments serve to perturb the initial conditions of a moral situation until one's intuitions vary. This can be philosophically useful insofar as we may be able to analyse {\it which} salient feature of the dilemma caused that variance.

    In the case of machine ethics, however, we have seen that moral dilemmas like the trolley problem are used as a {\it validation proxy} so that `if the implementation can resolve a dilemma in a particular manner, then it is deemed to be a successful implementation of ethics in the robot/software agent' \citep[2382]{Nallur-2020}. This is a category mistake. Moral dilemmas do not tell us what the truth is about whether a particular action is ethical or not; rather, they serve to create new avenues for inquiry. Regardless of one's metaphilosophical views concerning the ultimate purpose of thought experiments, moral dilemmas have no right answer by design. To suggest that agreement on ethical decision-making in trolley problems is {\it prior} to moral theorising about AI application presupposes that we have already settled important meta-ethical questions.\footnote{For example, if it were determined that a utility calculus is the `correct' normative theory, then we {\it could} use moral dilemmas as a validation tool. However, no such determination has been made.}

\section{Some Meta-Ethical Considerations}

    An anti-realist about ethics may, at this point, protest that there are no objective matters of fact about ethics. 
    Therefore, using human data from decisions in moral dilemmas as a benchmark for AI systems is certainly the closest we can get to measuring ethical behaviour---namely, maximising social acceptability. So, the Moral Machine Experiment data is the correct tool for this job.
    
    This is true, but it is also beside the point. Although some authors appear to be sensitive to the targets of their benchmark---i.e., the extent to which, all things considered, a human would accept the decision that an AI system made---it is much more common for there to be a significant conceptual gap between perceived targets and the actual targets of this research. What is problematic here is that researchers often appear to imagine that they are getting at one thing (`facts' of ethics) when they are really getting at another (sociological facts). It is perceived and therefore presented as though it is the former. This constitutes a derangement of the concept by which, over time, it comes to stand in for the thing itself---this will be a problem as we advance, for all the same reasons that any algorithmic bias is a serious social and philosophical problem.

    \section{Conclusion}

    Moral dilemmas are used in machine learning to provide circumstances where no ethical option is available to the (artificial) agent making the decision. However, to say that we want an autonomous system to minimise the unethical outcomes under these circumstances presupposes that we already know what the unethical outcomes to be minimised are---i.e., that we have already sorted out the relevant meta-ethical questions.%
        \footnote{Note that in response to the problem of enabling autonomous systems to distinguish between available choices and to choose the `least unethical' one, \citet{Dennis-et-al-2016} suggest that the pressing question to be resolved is `how can we constrain the unethical actions of autonomous systems but allow them to make justifiably unethical choices under certain circumstances?' But this presupposes that we already know what it means for a decision to be the least unethical. As is common, \citet{Dennis-et-al-2016} seem to understand `least unethical' in terms of `least unacceptable' by the standards of some subset of society.} 
    For example, utilitarianism might give a straightforward (and occasionally morally repugnant) answer when deciding between individuals and groups; however, it is less obvious how to calculate what potential future expected utility of a doctor's life will generate that a criminal's life will not. Never mind the fact that it is fallacious to suppose that because most people do reason this way, AI systems ought to reason this way; even if such a calculation is possible, it will always be relative to some frame---increased utility for whom?

    Using trolley-style problems in the context of autonomous vehicles as a case study, I have argued that researchers engaged in projects seeking to benchmark ethics are not measuring what they take themselves to be measuring. As we have seen, moral dilemmas are taken to provide something like a ground truth against which an algorithm can be benchmarked. But, this approach to ethical AI systems fails to appreciate the purpose of philosophical thought experiments in the first place. Lack of awareness of this fact sets a dangerous precedent for work in AI ethics, because these views get mutually reinforced within the field, leading to a negative feedback loop. The actual target(s) of AI ethics, by dint of being in the realm of moral philosophy, are already highly opaque. The more entrenched the approach of benchmarking ethics using moral dilemmas becomes, as a community-accepted standard, the less clearly individual researchers will see how and why it fails. 
    
    This also sets a dangerous precedent when we consider that the majority of AI research is now being done `in industry' (for profit) rather than in academia. Suppose a community-accepted standard for calling, e.g., an autonomous vehicle `ethical' is that it performs well on a set of trolley-style problems, which have been entrenched within the research community as an acceptable benchmark. As noted above, what is actually being measured is how well the machine accords with some set of humans on average, not how ethical the machine actually is---relative to some meta-ethical standard.

    This is not to say that moral dilemmas are never inappropriate in the context of AI systems. However, as with any system that uses proxies for measuring alignment of objectives, rather than the objectives themselves, it will be increasingly important that (1) the proxies used are actually representative of the true target, and (2) researchers are aware of what they are actually measuring. Of course, much more work needs to be done in the field of machine ethics to understand the relevant proxies for moral decision-making.  Even so, it should be clear that using trolley-style problems (or moral dilemmas more generally) as an elucidatory tool is neither prior to, nor follows from, moral theorising about AI applications.

\section*{Conflict of Interest}

On behalf of all authors, the corresponding author states that there is no conflict of interest.

\singlespacing
\small
\bibliographystyle{apalikelike}
\bibliography{Biblio}

\end{document}